# Erbium location into AlN films as probed by spatial resolution experimental techniques

V. Brien[*] and P. Boulet

Institut Jean Lamour, UMR 7198, CNRS, Université de Lorraine, Boulevard des Aiguillettes

B.P. 239, 54506 Vandœuvre-lès-Nancy Cedex, France.

## Abstract

This paper presents a thorough experimental investigation of erbium-doped aluminium nitride thin films prepared by R.F. magnetron sputtering, coupling Scanning Transmission Electron Microscopy X-ray-mapping imagery, conventional Transmission Electron Microscopy and X-ray diffraction. The study is an attempt of precise localisation of the rare earth atoms inside the films and in the hexagonal würtzite unit cell. The study shows that AlN:Er$_x$ is a solid solution even when x reaches 6 at.%, and does not lead to the precipitation of erbium rich phases. The X-ray diffraction measurements completed by simulation show that the main location of erbium in the AlN würtzite is the metal substitution site on the whole range. They also show that octahedral and tetrahedral sites of the würtzite do welcome Er ions over the [1.6-6 %] range. The XRD deductions allow some interpretations on the theoretical mechanisms of the photoluminescence mechanisms and more specifically on their concentration quenching.

---

[*] Corresponding author. Tel : +33 (0)383684928. E-mail address : valerie.brien@univ-lorraine.fr

# 1. Introduction

Rare earths (RE) activated luminescent materials are largely used in many domains as wide as telecom, medical applications, radiation and light detection, data recording and can be used to produce luminescent lamps, solid-state lasers or optical displays. Lanthanide atoms on their own are not natural emitters for the intra 4f energetic transitions are parity forbidden. However, when adequately inserted in a semiconductor or some insulators, the trivalent RE ions can be rendered optically active. The relaxation of the selection rules due to crystal field effects can lead the ions to produce sharp and strong luminescence peaks. Although the wavelength of emission does not change when changing the hosting matrix thanks to the electronic protection screen formed by the $5s^2$ and $5p^6$ orbitals, the intensity of emission is matrix dependant, and is at the source of a lot of research works. One can refer to the two general works by Kenyon [1] or by Liu [2] or find some examples in [3] or [4].

Among these studies, erbium attracts special attention due to its application field and to the high density of works present in the literature. Indeed, the aluminosilicate glasses composing optical fibres, need to have sources and amplifiers working at wavelengths that are compatible with the window of information transmission which costs least possible energy, that is coincides with the low-loss area of the absorption spectrum. The fortunate infra-red emission of erbium at 1.535 μm resulting of an electronic transition between the $^4I_{13/2}$ and the $^4I_{15/2}$ levels, makes the rare earth compatible for such requirements and explains why many studies address the incorporation of erbium in different matrices [5],[6],[7],[8],[9]...

First studied inside glasses, erbium has been incorporated in other materials than the ones constituting the fibers. Due to the possibility of easier integration of the emission optical sources, silicon and silicon based compounds (c-Si, nc-Si, a-Si, Si-SiO$_x$, SiGe, Si$_3$N$_4$, SiC, SiN) take a privilege portion of the works... [10],[11],[12],[13]. Researchers also tested and investigated the incorporation or erbium inside various oxides [14],[15],[9] or into III-V and II-VI semiconductors (GaN, AlN, ZnS, ZnO). The temperature quenching of the optical centres emission is known to be smaller in wide band gap semiconductors. This is why, AlN, with its theoretical 6,2 eV band gap, is one of the most interesting matrices to study the incorporation of erbium and other rare earths emitting in the whole visible spectrum up to the ultra-violet region [17],[18],[19],[20],[21],[22],[23].

The theoretical mechanism of optical emission of the trivalent rare earths ions in these matrices is not fully understood. The different authors of the community however agree that the localisation of the rare earth inside the solid, notably the symmetry of the site and the crystal field strength is a key question to understand the photoluminescence mechanism and optimise the emissive properties for technological applications [1], [24], [24], [25]. The symmetry of the site actually rules the way the electronic ions levels are split and can change completely the excitation-emission optical mechanism, resulting in variable intensities of the optical centres emissions. As underlined by Penn in a study on the electrical pumping of the GaN:Eu system for laser applications: each incorporation site possesses its own excitation mechanism [26].

From the point of view of the desired requirements for industrial applications, one of the most restricting parameters to increase the intensity is the quantity of ions one can incorporate inside the matrix. Many authors show that beyond a threshold concentration of the doping element inside the matrix, the emission mechanism loses its efficiency; the authors speak of "concentration quenching" [27], [28], [29]… This phenomenon is usually interpreted by two different processes: (1) the loss of energy is due to $RE^{3+}$-$RE^{3+}$ electron transfers: the potentially emitter ion (luminescent centre) is unexcited and it reduces the lifetime of the emission state limiting so the radiative emission ratio or leading to non radiative decay or/and (2) the fall is due to the precipitation of erbium rich aggregates, because the limit of solubility of the rare earth inside the welcoming material has been reached.

Another key point for the mechanism is the role of crystallographic defects (point defects, or grain boundaries) and their proximity to the emission centres [30][23].

As a consequence, knowing where the ions are located inside the lattice or inside the grains is of very high interest to try to understand the optical emission mechanisms.

In this work, we present a morphological and structural study of AlN:Er samples containing variable erbium contents. Observations were realised both in direct space (Transmission Electron Microscopy -TEM- and Scanning Transmission Electron Microscopy -STEM- X-ray mapping imagery) and in Fourier space (by X-ray or electron diffraction). The erbium doped aluminium nitride films were prepared by reactive radio frequency magnetron sputtering at room temperature. The films, containing different Er concentrations were earlier studied by steady state photoluminescence, photoluminescence excitation spectroscopy and time-resolved photoluminescence in a previous work [23]. The anterior work confirmed the classical behaviour of concentration quenching of these kinds of samples, by showing the existence of an optimal value for the infra-red photoluminescence signal for an erbium concentration at 1 at.%. Our goal here was then to try to get information on the precise localisation of the erbium cations throughout the samples, or at least suggest stronger hypotheses than the ones usually found in the literature on the location of the ions inside the aluminium nitride matrix. We also wanted to know if as suggested by some authors the concentration quenching in III-V, unlike some other systems, is due to the fact that the rare earths reach their solubility limit [31],[32].

## 2. Experimental procedure

The AlN:Er coatings were deposited in a home-made unbalanced reactive radio-frequency magnetron sputtering system. No intended heating on the substrates was used during deposition. A 3 mm thick aluminium target (99.99% purity) ~~with~~ 60 mm in diameter was used as the sputtering source, which was fixed on a water cooling copper back plate. Erbium was introduced by co-deposition using different specific targets containing Er sectors with adequate surface. The target-sample distance is equal to 150 mm. Substrates used are silicon (1 0 0) with dimensions of 200 mm × 200 mm × 1 mm. Prior to deposition, all substrates were ultrasonically rinsed sequentially in acetone, ethanol, and deionised water for 10 min. The base pressure of the chamber before deposition was $1 \times 10^{-6}$ Pa, and the deposition pressure 0.5 Pa was reached with a constant flow of argon and nitrogen gas mixture regulated by mass flow controllers. The $N_2/(Ar + N_2)$ percentage in the gas mixture was set to 50 %. Prior to the film deposition, 15 min of pre-sputtering using an Ar plasma was performed to clean the target surface. The sputtering power $P_m$ was set to 300 W and no thermal annealing treatment was performed after deposition. The control of the thickness is achieved by real time optical interferometry. Samples thickness is equal to 500 nm. Erbium contents in samples varied from 0 to 6 at.%. The erbium concentration was determined by Electron Dispersive Spectroscopy of X-rays (EDSX) and by the highly sensitive Rutherford Backscattering Spectrometry (RBS) technique. More precision on EDSX chemical calibration method could be found in Brien et al [33].

The nanostructural evolutions of the coatings were characterised with a CM20 or a CM200 PHILIPS transmission electron microscope, both running at an accelerating voltage of 200 kV. Most samples were prepared by microcleavage only one sample was thinned down by the Focused Ion Beam (FIB) technique.

The X-ray diffraction measurements have been performed using a Bruker D8-Discover diffractometer equipped with an Eulerian cradle. All the measurement have been performed with the CoK$\alpha$ wavelength ($\lambda$ =0.179026 nm) with a point focus of 1mm. The detection was assured either by a LynxEye Detector covering 3° in 2$\theta$ for the de-texturation analyses. The detexturation measurement is performed by measuring the sample at different $\chi$ angle with a step size of 1° in $\chi$. The sample is placed on fast spinner allowing the sample to rotate perpendicular to the diffusion plan. The "detextured" pattern is obtained by summing all the recorded patterns along $\chi$, after having corrected the intensity due to instrumental defocusing effect obtained from a textured free AlN sample.

## 3. Results

### 3.1 TEM investigation

After the PL characterisation of the samples showing the drop of the infra-red PL signal beyond 1 at.% on the 0 - 3.6 % range, we prepared on purpose a sample with 6 at.% of erbium in order to push the effects of erbium incorporation inside the würtzite, and study it. All the films (with erbium content ranging from 0 to 6 at. %) were sampled by microcleavage for quick TEM observation. The crystalline morphologies of the films are in these experimental conditions all polycrystalline and made of columnar grains whose width is around a few tenths of nanometers. This was extensively studied by the team in previously published work and has been presented in references [34], [35] and [36]. The adopted morphologies are fully consistent with what could be expected after reading the Thornton *"Structure Zone Diagram"* except that the present films are dense and exhibit no porosity. The stability of the crystalline morphology by this deposition process when doping the films with Erbium rare earth was presented in references [23] and [33]. As elemental X-ray mapping by Scanning Transmission Electron Microscopy with an X-ray energy dispersive spectrometer usually allows to get good indications in analysing nano-scale features in materials such as fine precipitates and interfaces/boundaries, we recorded X-ray mappings on the richest sample in erbium. It should be noted in passing that the average oxygen content was found similar for all samples within the range 5 - 8 at.%. Fig. 1 shows the four maps for the Al, N, O and Er elements using the adequate threshold energies of X-rays (cf. Fig. 1). Unfortunately, the spatial resolution on our samples was not high enough. Indeed, the thickness of the specimen (film thickness = 500 nm) degraded the spatial resolution of the mapping (from an optimal instrumental value of a few nanometers to around 10 nm here) and limited the observations. However, the analysis of the erbium map (Fig. 1d) allows to conclude that no precipitate of size above 15 nm is present in the 6 % doped AlN:Er sample.

Based on that, we recorded the electron diffraction patterns of the latter sample to try to detect the signature of either erbium itself or a phase rich in erbium. We selected a big area to maximise the chance to detect such phases (Fig. 2a). To compare the doped sample with the undoped one, the electron diffraction pattern recorded on the AlN sample prepared without erbium has been placed nearby (Fig. 2b). One can see that the two patterns contain the same amount of rings and that they are localised at the same distance from the transmitted spot. To ensure that weak intensity would be revealed (corresponding for instance to a small ratio of a new phase), longer exposure times were used on the doped sample and patterns were taken on different bits sampled on the specimen, along either cross or top views. The results are univocal. All patterns look alike: absolutely no other ring or spot different from the ones of ~~the~~ AlN würtzite could be detected on the AlN:Er (6%). As landmarks, reference guide rings using the AlN würtzite phase (JCPDS-file Nr 25-1133) and using the most probable pure erbium phase (pure erbium most common form is a hexagonal phase: JCPDS-file Nr 2-0930) have been drawn on the pattern for comparison. The same result was obtained on the 3.6 % erbium content AlN:Er film.

To reinforce the conclusion of this result, if it was necessary, one has recorded dark field images on different samples by selecting with the smallest aperture of the microscope an angular portion of the AlN rings ($\{0002\}$ and $\{10\bar{1}1\}$ of the diffraction pattern given in Fig. 2b). Special care was taken to make sure that the first AlN ring was avoided by judiciously choosing an adequate portion (as exemplified by P in Fig. 2a) of the diffraction rings as the AlN diffraction first ring theoretically superimposes to the first erbium ring (see theoretical rings in Fig. 2 and respective rings placed after the parameters from the JCPDS files). The strategy is here to try to evidence nodules or precipitates that would possibly not represented enough volume to produce a recordable intensity in diffraction, and that would be present for example at the grain boundaries of the columns of aluminium nitride, and would appear by leaving dark contrast at the edges of the white columnar grains. These types of images were recorded on microcleaved samples (0 %, 3.6 % and 6%) (cf. Fig. 3) and on a FIB thinned sample (1 %) (cf. Fig. 4). For that purpose, we prepared a specific 100 nm thick AlN:Er film doped with 1 at.% of erbium. It was impossible to detect the presence of such aggregates on these images, whatever the level of doping and even for the samples doped at 6 at.%. If such precipitates exist, their size would be under the spatial resolution of the images. The resolution on microcleaved samples can be estimated to 3 nm (the instrumental Philips CM200 microscope resolution is deteriorated by the specimen thickness). This is why a sample was thinned down by using FIB to reach the resolution of 0.5 nm. As spotted by the different arrows on the images displayed in Fig. 4 the columnar grains exhibit straight edges.

So, TEM investigation (either STEM-EDSX, electron diffraction or dark field imaging) did not allow to detect any presence of erbium precipitates (at least not under 0.5 nm for the 1 % doped film and not under 3 nm for the 6 % doped AlN:Er film), and this even when the AlN:Er samples were containing 6 at.% of erbium. If erbium did not feed any nucleates of any phase, one can suppose the doping atoms entered the würtzite lattice. Electron diffraction is not the ideal technique to get accurate measurements on lattice parameters. We have investigated the Fourier space by X-ray diffraction.

*3.2 X-ray diffraction investigation*

As many AlN films prepared by magnetron sputtering, the aluminium nitride films doped with erbium and prepared for this study are polycrystalline and textured. In order to get a precise pattern of the AlN film, $\theta/\theta$ X-ray patterns at different $\chi$ (tilt angle) have been recorded as explained in the experimental paragraph. Fig. 5 displays the intensity variation of XRD pattern versus the $2\theta$ Bragg angle and the tilt angle $\chi$ of the AlN:Er$_x$ sample (x =1 at.%). All diffraction peaks of AlN hexagonal structure are observed with the sequence $(10\bar{1}0)$, $(0002)$, $(10\bar{1}1)$, $(10\bar{1}2)$ $(11\bar{2}0)$ and $(10\bar{1}3)$. Moreover we should notice here that no deviation of the peak position is observed with the orientation of the film. This indicates that here the film is not under stress due to the substrate interaction. The evolution of the intensities versus $\chi$ angle of the diffraction

Bragg peaks indicates a texture of the film. These observations have been done for all the samples ranging from 0 to 6 at.% of erbium.

In order to gain a greater insight into the structural properties of these films; know if precipitated phases could be recorded, or follow the evolution of the cell parameter: cumulated X-ray patterns were recorded as a function of erbium content (presented in Fig. 6). These patterns were obtained by summing all the patterns obtained versus $\chi$ angle. As the sample is mounted in a fast spinner, the result is a "detexturized" pattern, with the proper intensities ratio between the diffraction peaks. As shown in Fig. 6, these patterns only exhibit the peaks of AlN würtzite type, plus the silicon peaks coming from first harmonic of the (004) silicon peaks of the substrate. No erbium rich phase could be traced here, even inside the AlN:Er films doped with 6 at.% of erbium. The unit cell parameters a and c of the hexagonal unit cell were achieved by using EVA software and for good precision of results, were made by taking account of the diffraction peaks at highest angle region (not shown in Fig.5 or 6) for each sample. Fig. 7 displays the graph of the obtained values as a function of the erbium content measured in the films. These two plots (Fig. 7) show that the evolution of the lattice parameter a is linear and follows a Vegard's law whereas the c lattice shows a step-like at 1% and follows a linear increase from 1.6%. So, the XRD investigation performed here by detexturation on the samples from 0 to 6 at.% of erbium in the AlN films proves that the rare earths are incorporated in the matrix and that they participate to the lattice even at 6%. This result is consistent with the TEM results presented above. These results demonstrate that AlN:Er$_x$ is a solid solution on the entire studied range x = 0 − 6%. No crystallographic saturation like the one observed by Chen et al. in [37] in the Er doped GaN lattice occurs here. The rare earth erbium atom entirely participates to the crystallographic lattice, and does progressively expand the würtzite lattice due to its radius larger than the Al one (covalent radii: 157 pm for Er, 130 pm for Al). The point of 1% doping appears special as it appears out of the linear law. This point will be discussed further.

By pushing the exploitation of the XRD results, one can also observe that all diffractograms exhibit the same number of peaks from 0 to 6%, proving this way that the crystallographic space group is invariant despite the introduction of an extra atom in the lattice. This is the proof that the doping rare earth atom adopts a crystallographic position respecting it: as a consequence a crystallographic insertion on sites, do not respecting the symmetry of the group P63mc has to be excluded. Now the assumptions on the erbium position have to consider 1/ the relative sizes of the atoms and the steric effect and 2/ the relative electronic affinity of atoms. To this regard the metal substitution site is the most probable. It is what is generally observed by channelling experiments when the rare earths are implanted on AlN and GaN films grown by MBE [37] or HVPE [38] or when co-deposited in AlN and GaN films grown by MOCVD[39][40] and is also consistent with the different available theoretical calculations performed on those systems (crystal field based [41],[39] or electron-phonon coupling based [42]). However, crystallographically speaking, tetrahedral and octahedral sites of the hexagonal lattice also have to be envisaged.

So to support this reasoning, we have simulated the evolution of the diffraction peaks intensities of the AlNEr würtzite structure by calculating the structure factor F$^2$ of the three main peaks as a function of the erbium content and assuming 100% of erbium atoms located on different crystallographic positions (we performed the calculation neglecting the effect of oxygen presence on diffracted intensities as 1/ oxygen diffusion factor is very close to the one of nitrogen and 2/ oxygen atoms are expected to occupy nitrogen sites). This calculation has been performed using the software FullProf [39]. Three different localisations have been calculated:

- The first localisation is in substitution of the Al atom i.e. at the Wyckoff position 2b (1/3,2/3,z=0.38).

- The second localisation is the tetrahedral site which corresponds to the position in between the aluminium and the nitrogen i.e. at the Wyckoff position 2b (1/3,2/3,z=0.69).

-The last position is the octahedral site which corresponds to the 2a Wyckoff position (0,0,z=¼).

The comparison of the $F^2$ structure factor simulation calculation and the experimental intensities of the X-ray peaks are presented in Fig. 8. They are presented on a graph displaying the experimental ratios $R_{hkl}^{exp} = I_{hkl} / I_{100}$ and the theoretical ratios $R_{hkl}^{th} = F_{hkl}^2 / F_{100}^2$ for hkl = 0 0 2 and 1 0 1 as a function of erbium content for the three postulates. Normalisation of the ratios was chosen to be done by dividing by the most intense peak of the würtzite 1 0 0. The intensities $I_{hkl}$ are the integrated surfaces of the corresponding XRD peaks.

The confrontation of the data is done by comparing the evolution of curves drawn by the points (i.e. the slopes) as it allows to overcome the absence of correction of some experimental parameters being sources of implicit systemic differences between the experimental records. The point 1% is discarded from the interpolation because of two reasons: 1/ an extra-dilatation of c lattice cell parameter is observed pushing the 1% point out of the Vegards'law and 2/ the ratios of XRD peaks $R_{002}^{exp}$ (1%) and $R_{101}^{exp}$ (1%) are also out of the linear trends. This singularity allows suspecting a different crystallographic behavior. It will be discussed further.

The evolution of experimental XRD peaks ratios as a function of the doping content is the best approached by the theoretical curve assuming a substitution of Aluminium by Erbium. Indeed slopes coefficients of the linear interpolating laws are ( $\dot{R}_{101}^{exp} = +0.07;\ \dot{R}_{002}^{exp} = -0.01$) as they are ($\dot{R}_{101}^{th} = +0.04;\ \dot{R}_{002}^{th} = +0.01$) for the substitution hypothesis, ( $\dot{R}_{101}^{th} = -0.12;\ \dot{R}_{002}^{th} = -0.04$ ) for the tetrahedral one and finally ($\dot{R}_{101}^{th} = +0.29;\ \dot{R}_{002}^{th} = +0.07$ ) for the octahedral hypothesis. However it is not sufficient. A strict comparison of the values leads to the conclusion that one needs to involve the two other types of sites. A combination of Al substitution and only one of the two other sites would not account for the positive and negative slopes of $\dot{R}_{101}^{exp}$ and $\dot{R}_{002}^{exp}$ respectively. Erbium would then mainly adopt a distribution on Al 2b sites and the remaining Er atoms, i.e. a very small amount, on both the octahedral and tetrahedral sites.

This can be consistent with the following crystallographic interpretation. One can imagine the rare earth is in pure substitution on the Al site until 1% (c increases by 0.4%) and that over, the lattice cannot put up with more expansion; more introduction of the big atom makes the rare earth to be located in the insertion lattice sites (octahedral and tetrahedral) which are now bigger due to the 1% point expansion: the lattice does not expand in the first place as the free space is filled ($\Delta$c/c (1.7%) = 0.4%); and further introduction of Er could then continue to fill the insertion sites on the 1.7 – 6% range pushing progressively the lattice parameter c according the Vegard's law.

The crystallographic considerations deducted here so far however bring elements to interpret the photoluminescence mechanisms of this system. The first point to take in account is the absence of extra erbium rich precipitated phases. $Er^{3+}$-$Er^{3+}$ interactions inside non optically active precipitates as put forward in some others works can then not be invoked here to justify the concentration quenching.

The XRD data presented here seem to show that until 1% the ions occupy exclusively the Al site ($C_{3v}$ symmetry) which is an optimal crystallographic environment for efficient PL. Above this value, the centres that were active at 1% are hindered as PL efficiency drops.

Such an observation implies that (1) the Er ions located in insertion locations are not optically active and that (2) the extra erbium ions are detrimental to the optical production of the optically active centres located on the Al sites. Such effect could be explained by $Er^{3+}$ - $Er^{3+}$ interactions (like co-operative up-conversion, energy migration or cross-relaxation) which are indeed known to occur when close enough, that leads to non-radiative decay.

As XRD data showed that the lanthanide ions are disseminated throughout the volume of the grains all over the [0 - 6%] range, an average statistical physical distance between $Er^{3+}$ ions can be calculated for the 1 % doped nitride assuming the ions distribution throughout cells is purely homogeneous and random in ~~the~~ volume. The calculation can either be done by taking the density of AlN, ($d$ = 3.28) or deducing it from crystallo-chemical considerations taking the number of atoms in a lattice and its theoretical volume. The distance obtained is 10 Å, that is 2 c or around 3 a (one recalls the cell parameters: $a_{AlN}$ = 3.11 Å and $c_{AlN}$ = 4.98 Å). So, the crystal statistically possesses then one lanthanide every two cells along c, whereas every three cells along a. Beyond 1 % the supplementary ions start to fill the insertion sites randomly, diminishing in parallel the average distance between 2 Er ions. So PL efficiency seems to be hindered by a too high density of centres. Under an average distance of $d_0$ = 10 Å between $Er^{3+}$ ions, the incorporation itself or the crystallographic perturbations (vacancy type defects) going with it would be close enough to modify the radiative emission from the RE shell transitions. The filling of an octahedral or tetrahedral site may be accompanied by the creation of vacancies. The second atomic shell of erbium ions located in insertion sites should also present oxygen atoms. It is not straightforward to deduce further. The situation is complex. Both aluminium or nitrogen vacancy type defects (or larger) can change the local structure around RE atoms, and could then affect the transition probability modifying the way the emission rate of the intra-4f shell transitions from RE ions. On the other way, oxygen presence is also known to have a great influence on the optical mechanisms. The drop is also fed by the increase of the probability of occurrence of adjacent sites (metallic one and insertion ones) which is real above 1.7%.

The theoretical calculation of luminescence efficiency of activator ions as a function of the activator concentration; recognising that only activator ions not having other activators on the adjacent site are capable of luminescence and assuming a random distribution; has already been achieved on other luminescent systems (Mn in $ZnF_2$) [40]. This theoretical curve reflects quite the shape of the experimental curve of integrated photoluminescence as a function of the erbium content that could be measured on the present samples (published in [23]). Indeed, in both cases the curves increase until a maximum; which is followed by a drop. So, a random distribution throughout the grains appears to be sufficient to account for the presence of a maximum in the photoluminescence efficiency and for the concentration quenching.

Another hypothesis that could be stated at this stage, would not consider the only distance criteria between Er sites as being sufficient to promote loss of energy by $Er^{3+}$ - $Er^{3+}$ interactions, but would take account of the nature of the symmetry of the new Er ions, and would take for responsible the degeneracy of levels of higher symmetry (symmetries of octahedral and tetrahedral sites are superior to the one of the metallic site in würtzite).

Another hypothesis that could be stated at this stage, would not consider the only distance criteria between Er sites as being sufficient to promote loss of energy by Er3+-

Er–Er3+ interactions, but would take account for of the nature of the symmetry of the new Er ions, and would take be responsible for responsiblethe degeneracy of levels of higher symmetry (symmetries of octahedral and tetrahedral sites are superior to the one of on the metallic site in würtzite).

Conclusion

TEM and XRD analyses were performed on Er doped sputtered AlN samples with variable erbium content [0 – 6 at. %]. [0–6 at.%]. The absence of extra phases on dark field, STEM images or electron and X diffraction data on the whole range proves the incorporation of the rare earth inside the AlN würtzite. Erbium is shown to be a full part of the AlN hexagonal unit cell; it expands the lattice progressively and proportionally to its quantity: AlNErx is a solid solution until x = 6atomic % at.% whose lattice parameters follow aVegard's law. The solubility limit has then not been reached even at 6atomic %. at.%.

From the theoretical calculations of XRD intensities, one deduces that the lanthanide atom mainly occupies the regular substitutional metallic atom in the nitride and that from a certain amount (1.7 %) (1.7%) and part of it also fills the insertion sites of the hexagonal lattice (octahedral and tetrahedral sites). The rare earth ions adopt a random volume distribution inside AlN grains. The phase can then be written Al1-x1−xNOzErxEry.

From these deductions, a discussion was made to deepen the understanding of the photoluminescence concentration quenching measured on these samples in previously published works. This study brings theproof that the limit of solubility (or presence of extra erbium rich phases) is not the limit to efficient photoluminescence in this kind these kinds of samples and cannot be taken responsible for the concentration quenching effect.

The issue of the physical mechanism ruling the PL quenching was discussed and some assumptions are made. The work still leaves us to further debates debate on whether the concentration of PL quenching is related to Er3+- Er–Er3+ interactions that occur either because the Er3+- Er–Er3+ distances are under a threshold or because of the degeneracy of the erbium levels of the octahedral and tetrahedral sites due to higher symmetry than the metallic site or due to oxygen electronegativity effect on erbium orbitals. The distance considerations lead to an optimum distance of 10 Å as the ideal distance between two Er3+ ions to optimize optimise the photoluminescence mechanism.

The role of site environment chosen by the erbium ions in the PL mechanism still has to be pushed. At that stage a dedicated vacancy and/or spectroscopic studies on these samples would indeed be very fruitful to deepen the comprehension of the optical mechanisms. The authors envisage completing the study by performing spectroscopy targeting the edge energy of the rare earth.

Conclusion

TEM and XRD analyses were performed on Er doped sputtered AlN samples with variable erbium content [0 – 6 at. %]. The absence of extra phases on dark field, STEM images or electron and X diffraction data on the whole range proves the incorporation of the rare earth inside the AlN würtzite. Erbium is shown to be a full part of the AlN hexagonal unit cell; it expands the lattice progressively and proportionally to its quantity: AlNEr$_x$ is a solid solution until x = 6 at.% whose lattice parameters follow a Vegard's law. The solubility limit has then not been reached even at 6 at.%.

From the theoretical calculations of XRD intensities, one deduces that the lanthanide atom mainly occupies the regular substitutional metallic atom in the nitride and that from a certain amount (1.7 %) part of it also fills the insertion sites of the hexagonal lattice (octahedral and tetrahedral sites). The rare earth ions adopt a random volume distribution inside AlN grains. The phase can then be written $Al_{1-x}NO_zEr_xEr_y$.

From these deductions, a discussion was made to deepen the understanding of the photoluminescence concentration quenching measured on these samples in previously published works. This study brings the proof that the limit of solubility (or presence of extra erbium rich phases) is not the limit to efficient photoluminescence in this kind of samples and cannot be taken responsible for the concentration quenching effect.

The issue of the physical mechanism ruling the PL quenching was discussed and some assumptions are made. The work still leaves to further debates whether the concentration PL quenching is related to $Er^{3+}$ - $Er^{3+}$ interactions that occur either because the $Er^{3+}$ - $Er^{3+}$ distances are under a threshold or because of the degeneracy of the erbium levels of the octahedral and tetrahedral sites due to higher symmetry than the metallic site or due to oxygen electronegativity effect on erbium orbitals. The distance considerations lead to an optimum distance of 10 Å as the ideal distance between two $Er^{3+}$ ions to optimize the photoluminescence mechanism.

The role of site environment chosen by the erbium ions in the PL mechanism still has to be pushed. At that stage a dedicated vacancy and/or spectroscopic studies on these samples would indeed be very fruitful to deepen the comprehension of the optical mechanisms. The authors envisage completing the study by performing spectroscopy targeting the edge energy of the rare earth.

Figure captions

Fig. 1. X-ray microanalytical mapping on a cross section AlN:Er doped with 6 at.% of erbium using the energy values of a/ Al, b/ N, c/ O and d/ Er. Magnetron R.F. deposited films were prepared by using a working pressure of 0.5 Pa and a power of 300W. No coarsening of erbium rich phase can be evidenced here.

Fig. 2. TEM diffraction patterns obtained a/ on the microcleaved AlN:Er film with 6 at.% of erbium b/ on the microcleaved non doped AlN film. Lattice constants found in the reference database JCPDS files were used to draw the theoretical signatures of AlN würzite (green rings: JCPDS file Nr 2-0930) and most probable pure erbium phases (red rings: JCPDS file Nr 2-0930). P = Exemple of selected area to build images of Fig. 3 and Fig. 4.

Fig. 3. TEM Dark Field images recorded by using the {0002} and {10$\bar{1}$1} rings of würzite on cross sections of **microcleaved** samples a/ AlN:Er Erbium content = 6 %, b/ AlN:Er Erbium content = 3.6 % and c/ AlN films. As indicated by arrows, no precipitation of any kind can be evidenced at the grain boundaries of the columnar grains.

Fig. 4. TEM Dark Field images recorded by using the {0002} and {10$\bar{1}$1} rings of aluminium nitride on cross sections of **FIB** thinned down on the AlN:Er sample (Erbium content = 1 %). No nodules can be evidenced at the grain boundaries of the columnar grains (arrows).

Fig. 5. XRD patterns of the 1 % AlN:Er sample using the Euler Circle diffractometer. Intensity is plot versus the tilt angle $\chi$ and the Bragg angle $2\theta$. Each diagram recorded with a step of 5° along $\chi$ are shown from 0 (bottom) to 75° (top). The three vertical lines are guides for the eye corresponding to the three Bragg Peaks (100), (002) and (101).

Fig. 6. Cumulated X-ray patterns showing the entire reciprocal space of AlN:Er films as a function of the erbium content.

Fig. 7. Unit cell parameters a and c obtained from the detextured patterns versus the erbium content. Experimental error bars are included in the point markers.

Fig. 8. Experimental (a) and theoretical (b) to (d) ratios $R_{hkl}^{exp\ or\ th}$ of the main XRD peaks of the AlN cell as a function of erbium content. For experimental data: $R_{002}^{exp}$ = I$_{002}$ / I$_{100}$, $R_{101}^{exp}$ = I$_{101}$ / I$_{100}$ and for theoretical $R_{002}^{th}$ = F$_{002}$$^2$ / F$_{100}$$^2$, $R_{101}^{th}$ = F$_{101}$$^2$ / F$_{100}$$^2$. 100% of Er atoms are supposed to: - (b) substitute to the aluminium atom - (c) occupy tetrahedral sites - (d) occupy octahedral sites.

Table 1. Theoretical intensities of structure factor $F^2$ of AlN:Er würtzite cell as a function of the crystallographic localisation of the rare earth inside the cell and versus the erbium content.

| Er % | 100% of Er atoms substitute Al atoms (1/3,2/3,z=0) | | | 100% of Er atoms are on the tetrahedral site (1/3,2/3,z=0.62) | | | 100% of Er atoms are on the octahedral site (0,0,z=¼) | | |
|---|---|---|---|---|---|---|---|---|---|
| | $I_{100}$ | $I_{002}$ | $I_{101}$ | $I_{100}$ | $I_{002}$ | $I_{101}$ | $I_{100}$ | $I_{002}$ | $I_{101}$ |
| 0,0 | 1000 | 622 | 957 | 1000 | 622 | 957 | 1000 | 622 | 957 |
| 1,0 | 1054 | 662 | 1056 | 1069 | 604 | 834 | 870 | 573 | 957 |
| 1,7 | 1093 | 692 | 1132 | 1118 | 592 | 753 | 784 | 541 | 957 |
| 2,8 | 1156 | 740 | 1253 | 1198 | 576 | 634 | 659 | 492 | 957 |
| 3,6 | 1203 | 776 | 1346 | 1257 | 566 | 555 | 574 | 459 | 957 |
| 6,0 | 1349 | 890 | 1650 | 1445 | 546 | 354 | 356 | 367 | 957 |

Fig. 1.

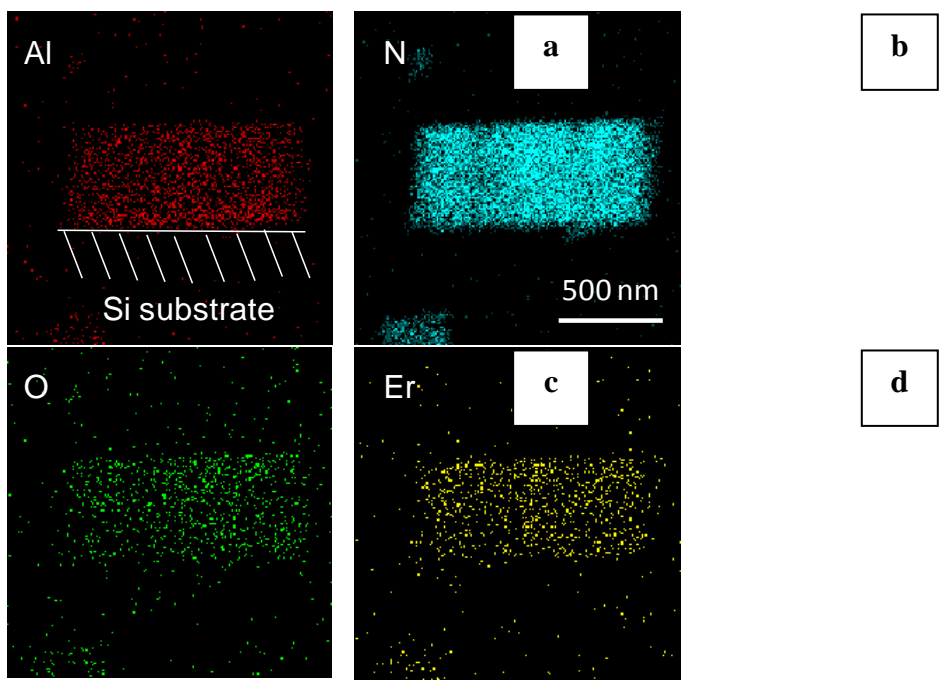

Fig. 2.

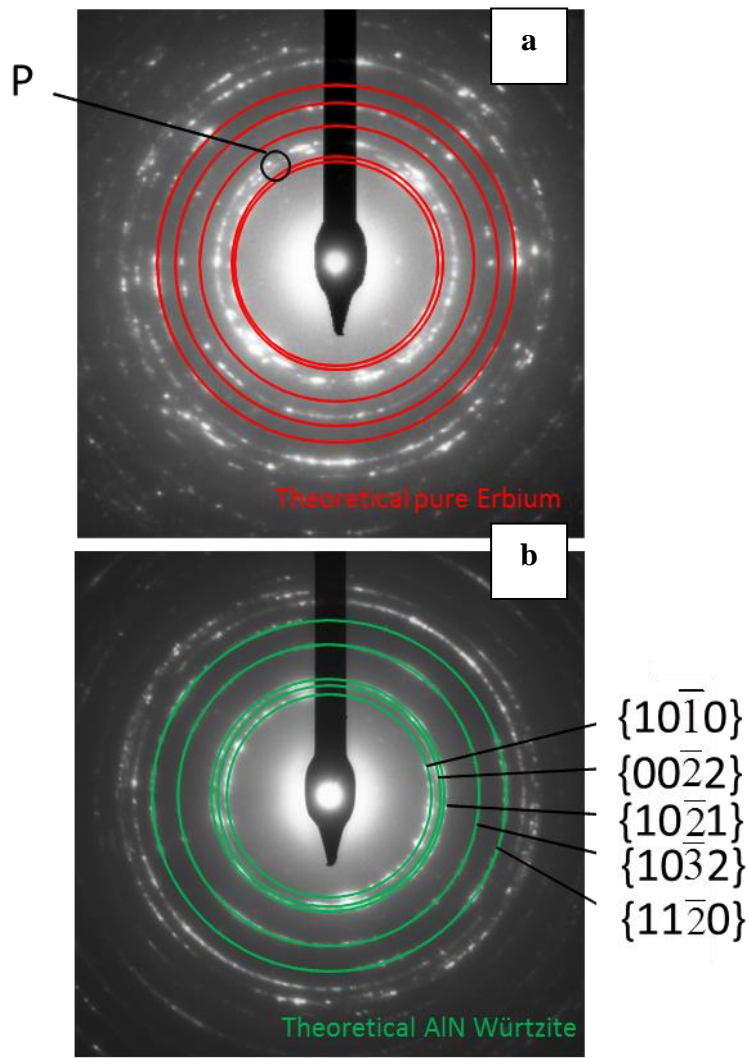

P

a

Theoretical pure Erbium

b

{10$\bar{1}$0}
{00$\bar{2}$2}
{10$\bar{2}$1}
{10$\bar{3}$2}
{11$\bar{2}$0}

Theoretical AlN Würtzite

Fig 3

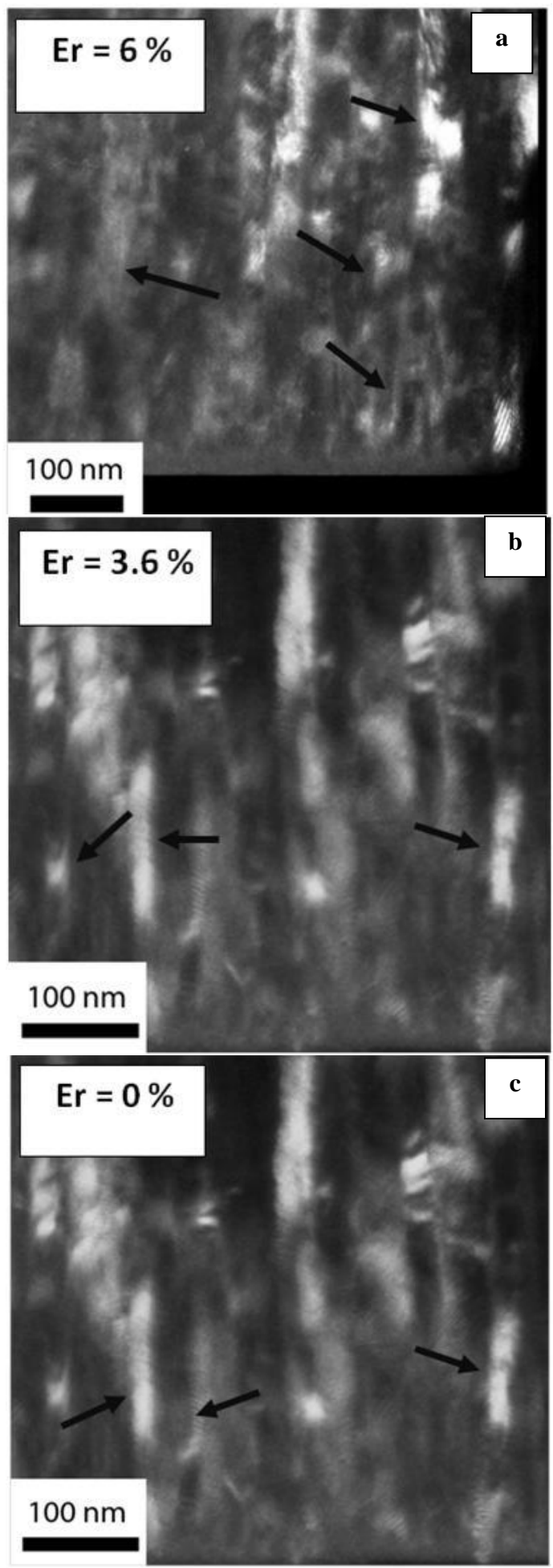

Er = 6 %

a

100 nm

Er = 3.6 %

b

100 nm

Er = 0 %

c

100 nm

Fig. 4.

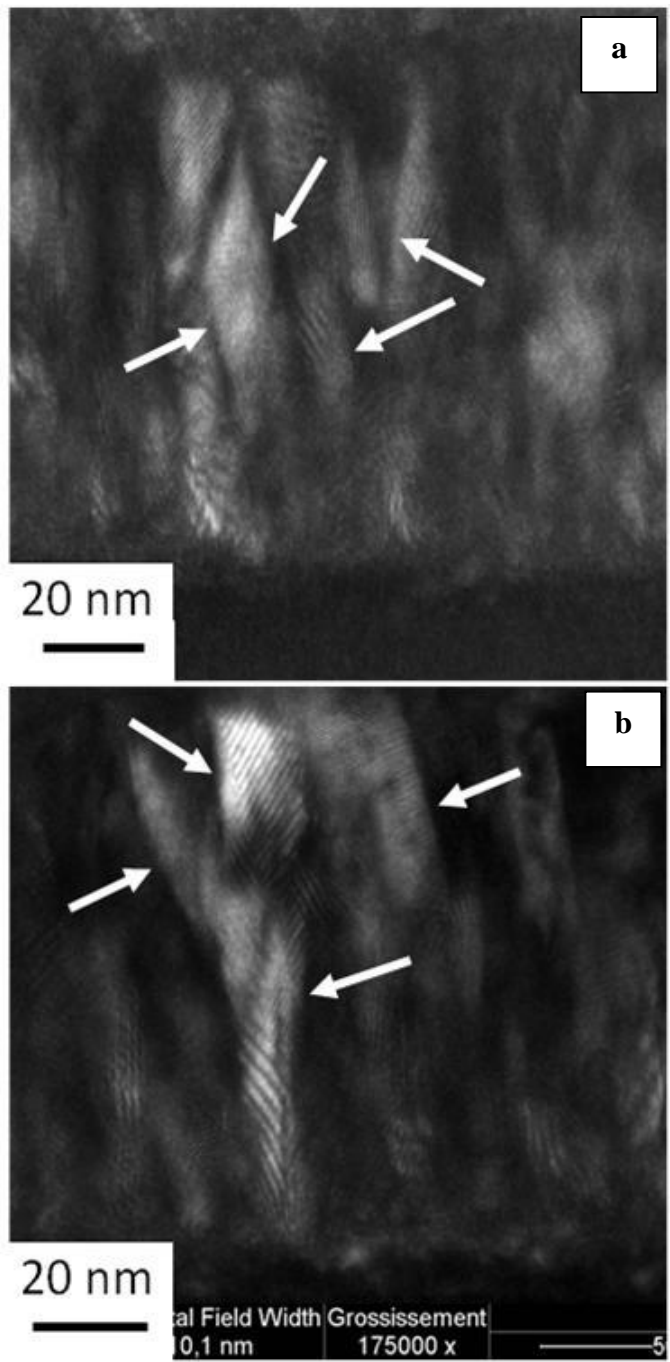

20 nm

20 nm

al Field Width | Grossissement
0,1 nm | 175000 x | 5

Fig 5

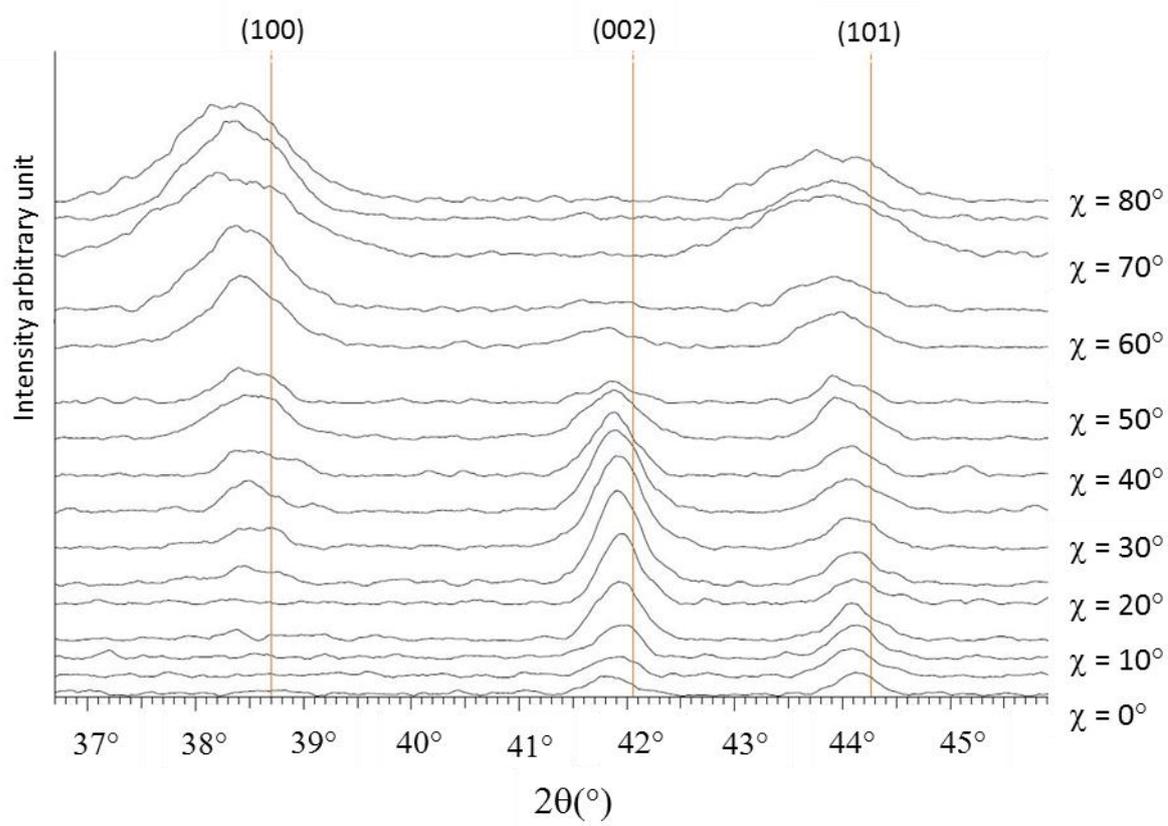

Fig. 6.

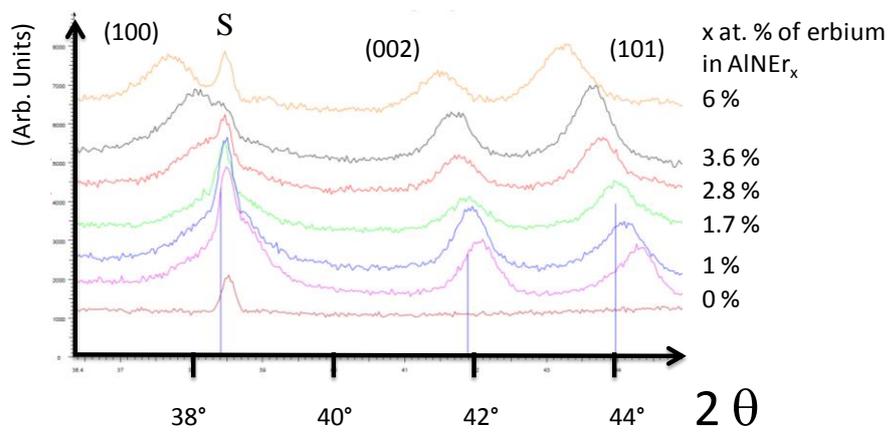

Fig. 7.

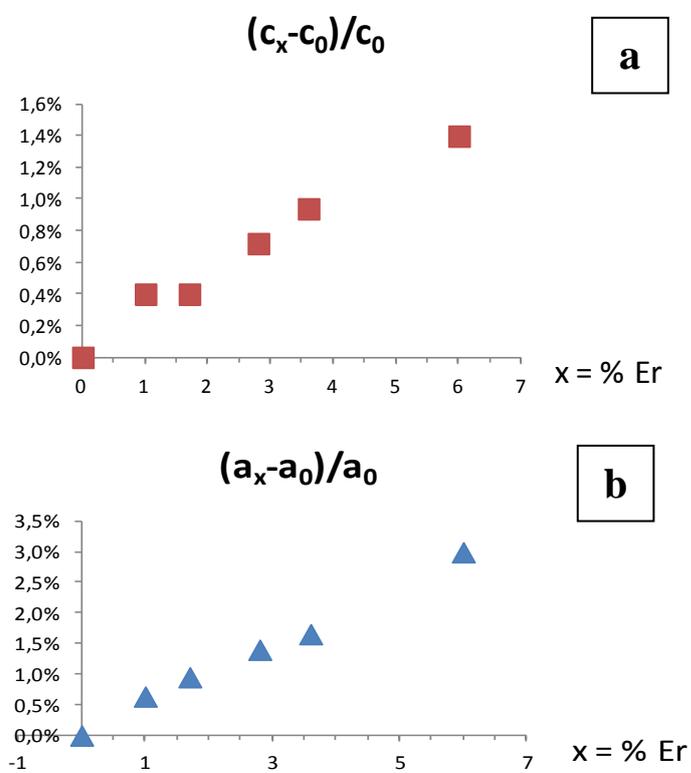

Fig. 8.

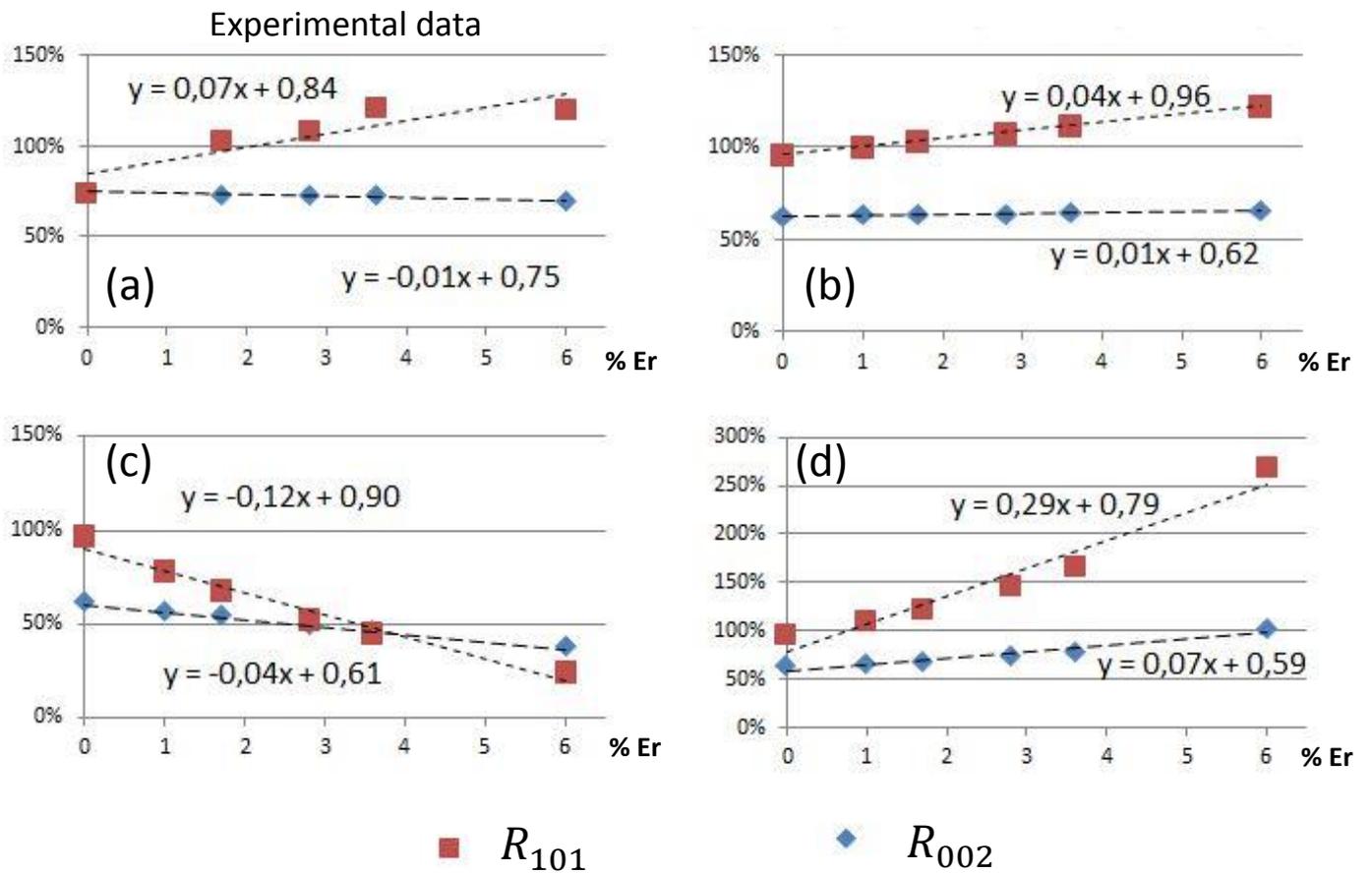

$R_{101}$        $R_{002}$